\documentclass[runningheads]{llncs}
\usepackage{graphicx}
\usepackage[caption=false]{subfig}
\usepackage[ruled,vlined]{algorithm2e}
\usepackage{hyperref}

\newcommand{\Sref}[1]{\S\ref{#1}}
\newcommand{\fref}[1]{figure~\ref{#1}}
\newcommand{\Fref}[1]{Figure~\ref{#1}}
\newcommand{\tref}[1]{table~\ref{#1}}

\begin{document}
\title{Climate Change Conspiracy Theories on Social Media}
\titlerunning{Climate Change Conspiracy Theories}
\author{Aman Tyagi\inst{1}\orcidID{0000-0002-6654-0670} \and
Kathleen M. Carley\inst{1,2}\orcidID{0000-0002-6356-0238} }
\authorrunning{Tyagi and Carley}
\institute{Engineering and Public Policy, Carnegie Mellon University, PA 15213, USA \and
Institute for Software Research, Carnegie Mellon University, PA 15213, USA
\email{amantyagi@cmu.edu, kathleen.carley@cs.cmu.edu}}

\maketitle              % typeset the header of the contribution

\begin{abstract}

One of the critical emerging challenges in climate change communication is the prevalence of conspiracy theories. This paper discusses some of the major conspiracy theories related to climate change found in a large Twitter corpus. We use a state-of-the-art stance detection method to find whether conspiracy theories are more popular among Disbelievers or Believers of climate change. We then analyze which conspiracy theory is more popular than the others and how popularity changes with climate change belief. We find that Disbelievers of climate change are overwhelmingly responsible for sharing messages with conspiracy theory-related keywords, and not all conspiracy theories are equally shared. Lastly, we discuss the implications of our findings for climate change communication.

\end{abstract}

\section{Introduction}

There is a virtually 100\% consensus among scientists that greenhouse gas emissions from human activity cause climate change \cite{doran2009examining}. Despite the overwhelming evidence, much public discourse shows open skepticism with many popular contrarian voices \cite{dunlap2016political,boykoff2004balance,painter2016climate}. In fact, it is believed that between 20\% to 40\% of the U.S. population considers climate change as a hoax or do not believe in its anthropogenic cause \cite{uscinski2017climate}.

Contrarian voices on climate change can be divided among different categories. For instance, there is a category of people who argue that climate change is real but is not caused by human activity. Another example would be people who believe that climate change is real, but they dispute the anthropogenic cause. However, the most alarming category are climate change deniers who outright reject climate science findings or the data as a hoax. Different ideologies drive most people who describe climate science findings and data as hoax \cite{uscinski2017climate}. One such facet of ideology is conspiratorial thinking. Previous studies have suggested that conspiratorial thinking is associated with beliefs about climate change. In other words, individuals who believe in conspiracies are more likely to refute the anthropogenic cause of climate change \cite{uscinski2017climate}. 

Conspiracy theories are ``unsubstantiated explanations of events or circumstances that accuse powerful malevolent groups of plotting in secret for their own benefit against the common good" \cite{uscinski2017climate}. People who believe in conspiracy theory might want to derive an explanation for any complex scientific fact from these theories. Conspiracy theories can be interlinked with each other, although they might not have any logical basis. In this paper, we would discuss some of the major conspiracy theories in climate change that are popular on a social media platform. These conspiracy theories present a significant challenge in removing the false narratives around climate change. Thus, it becomes essential to analyze these conspiracy theories. 

Previous work on climate change and conspiracy theories suggest that people believing in conspiracy theories are likely to believe that climate change is a hoax \cite{uscinski2017climate}. That work relied on the manual survey-based collection methods. Surveys are limited in finding nuanced beliefs and in studying extensive social network structures.  This paper uses extensive Twitter data to link beliefs about climate change and sharing of conspiracy related text. Thus, the main research question we answer is, \emph{In climate change discussion, do climate change Disbelievers share more conspiracy related terms compared to Believers?} To answer this research question, we scrape Twitter data for all the Tweets containing climate change and conspiracy related keywords. We then use a state-of-the-art stance detection method to find climate change Believers and Disbelievers \footnote{We define Believers as people who cognitively accept anthropogenic causes of climate change Disbelievers as those who reject the same.}. 

Moreover, conspiracies about climate change could be promulgated by bot-like accounts - automated user accounts - in addition to human actors. These bot-like accounts can further create confusion on well established climate change realities. Previous studies have suggested that ``bots seek to create false amplification of contentious issues with the intention to create discord" \cite{tyagi2020brims}. This paper examines whether or not bot-like accounts are more active in sharing conspiratorial messages in different belief groups. 

This paper begins by defining some of the common climate change-related conspiracy theories \Sref{sec:cons:conspiracies}. Next,  we discuss the method used to identify individual beliefs, keywords used to identify the conspiracy theories, and method used to find bot-like accounts \Sref{sec:cons:data-method}. We present our results in \Sref{sec:cons:results}. Our results (\Sref{sec:cons:results}) suggest that Disbelievers share most conspiracy theory related Tweets. Conspiracy theory related to chemtrails and geo-engineering is most popular in our dataset. However, conspiracy theory related to flat earth is most popular among Believers but rather used as a sarcasm. We also find that most Disbelievers share only one or two different conspiracy theories with climate change discussion. Finally, we discuss our findings and their implications in \Sref{sec:cons:discussion}.

\section{Major Conspiracies about Climate Change}
\label{sec:cons:conspiracies}

Conspiracy theories evolve with time and do not follow logical arguments. This section covers the most well-known conspiracy theories related to climate change and gives brief backgrounds about each. The list was created based on the author's findings, and readings of previous work on the same topic \cite{wikiconsipracy,tyagi2020brims,uscinski2017climate,tyagithesis}\footnote{List of example tweets can be found at \url{ https://github.com/amantyag/Climate_Change_Conpiracies/}}.

\begin{enumerate}

    \item Deep state: Followers of this conspiracy theory agree that there is a hidden government within the legitimately elected government that controls the state. Climate change is a hidden agenda of the deep state to further the deep's states motives. 
    \item Chem Trails: The condensation trails from the jet engines of an aircraft are erroneously recognized as consisting of chemical or biological agents. The theory posits that these trails are responsible for climate change. 
    \item Sunspots: Sunspots are a temporary phenomenon of reduced temperature on the Sun's surface \cite{ruzmaikin2001origin}. This theory asserts that sunspots and not human activity are causing climate change. 
    \item Directed Energy Weapon (DEW): A human-made weapon that damages its target by a highly focussed beam of energy. As per the proponents of this theory, the usage of DEWs is causing climate change. 
    \item Flat Earth: Advocates of this conspiracy theory do not believe that the earth is a sphere but rather believe that the earth is a flat disc. Climate is hence not governed by the standard scientific laws, and climate change is a hoax. 
    \item Geo Engineering: Enthusiasts of this conspiracy theory believe that governmental experiments cause climate change.  
    \item Unknown Planet: A ninth planet with a vast orbit and unknown to humanity is causing climate change. The effect of the planet will keep on increasing as it goes through its perigee.  
\end{enumerate}

\section{Data Collection and Method}
\label{sec:cons:data-method}

In this section, we first describe our data collection in \Sref{sec:cons:data-collect}. Second, in \Sref{sec:cons:method}, we describe our method to find climate change belief stance and the keywords used to find conspiracy theory related Tweets.

\subsection{Data Collection}
\label{sec:cons:data-collect}

We collected tweets using Twitter's standard API\footnote{\url{https://developer.Twitter.com/en/docs/tweets/search/overview/standard}} with keywords ``Climate Change'', ``\#ActOnClimate'', ``\#ClimateChange''. Our dataset was collected between August 26th, 2017 to September 14th, 2019. Due to server errors, the collection was paused from April 7th, 2018 to May 21st, 2018, and again from May 12th, 2019 to May 16th, 2019. We ignore these periods in our analysis. After deduplicating tweets, our dataset consisted of 38M unique tweets and retweets from 7M unique users.

\subsection{Method}
\label{sec:cons:method}

This section will first discuss the stance detection method used to identify climate change Believers and Disbelievers. Then, the keywords used to identify conspiracy related Tweets.

\paragraph{Stance Detection:} Labeling each user as a climate change Believer or a Disbeliever is a non-trivial task. The broader field of labeling users based on the position the user takes on a particular topic is called \textit{stance mining} \cite{mohammad2017stance}. We use state-of-the-art stance mining method which uses weak supervision to find Believers and Disbelievers \cite{kumar2020social}. The model uses text signals from Tweets along with retweet and hashtag network features using a co-training approach with label propagation \cite{zhu2002learning} and text classification. A set of seed hashtags are provided as a pro and anti stance signals to the model. The model then labels seed users based on the usage of these seed hashtags at the end of the tweet (endtags). The labeled and unlabeled users are then taken as input to the co-training algorithm. In each step, a combined user-retweet and user-hashtag network is used to propagate labels to unlabelled users. Concurrently, the text classifier uses the seed user's tweets to train an SVM \cite{cortes1995support} based text classifier to predict unlabeled users. A common set from text classification and label propagation of highly confident labels are then used as seed labels for the next iteration. The final classification is based on the prediction of the joint model using the combined confidence scores.\footnote{ We use the parameter values as defined in \cite{kumar2020social} as \{$k = 5000$, $p = 5000$, $\theta^{I} = 0.1$, $\theta^{U} = 0.0$, $\theta^{T} = 0.7$\}.} The model has been shown to be above 80\% accurate with multiple datasets.

We select hashtag \textit{\#ClimateHoax} and \textit{\#ClimateChangeIsNotReal} as Disbeliever seed hashtags and \textit{\#ClimateChangeIsReal} and \textit{\#SavetheEarth} as Believer seed hashtags. Hashtags \textit{ClimateHoax} has been shown to be used mostly by Disbelivers \cite{tyagi2020brims,tyagi2020affective}. We found similar results on using other Disbeliever hashtags reported in \cite{tyagi2020brims,tyagi2020affective}. We use \textit{ClimateChangeIsReal} and \textit{SavetheEarth} as Believer hashtags because of their semantics. Out of the 7M users, we classified 3.1M as disbelievers and 3.9M as believers \footnote{We randomly sampled 1000 users from each group to manually validate the results. We label a user as Disbeliever if we find any Tweet akin to someone who does not believe in climate change or anthropogenic cause of climate change. Otherwise, we label the user as Believer. We observe that the average precision from manual validation of 2000 users is 81.2\%.}.

\paragraph {Conspiracy Keywords:} We use the following keywords to identify if a Tweet is a conspiracy related Tweet. 
\begin{enumerate}
    \item Deep state: club of rome, clubofrome, clubrome, pizzagate, lizard people, lizardpeople, illuminati, deepstate, deep state, qanon
    \item Chem Trails: chemtrail, chem trail
    \item Sunspots: sunspot
    \item Directed Energy Weapon: dew, directed energy weapon, directedenergy
    \item Flat Earth: flat earth, flatearth
    \item Geo Engineering: geo engineering, geoengineering, weather modification, weathermodification
    \item Unknown Planet: planet x, niburu
\end{enumerate}

\paragraph{Bot Detection:} We label an account as bot-like or not using CMU’s Bot-Hunter \cite{beskow2018bot1,beskow2018bot2}. Bot-Hunter's output is a probability measure of bot-like behavior assigned to each account.
%Unless otherwise stated, we report our analysis for a probability threshold of 0.6 \footnote{We use 0.6 as this probability threshold gives us a lower false-positive rate than generally used 0.5.}. In other words, we classified an account as bot-like if output probability from Bot-Hunter was greater than 0.6. 

\section{Results}
\label{sec:cons:results}

Climate change Disbelievers share more conspiracy related Tweets than climate change Believers. We find that the number of Tweets and Retweets shared by climate change Believers (4830 and 3576) and Disbelievers (31084 and 14369) are an order of magnitude different. Disbelievers overwhelmingly share Tweets related to conspiracy theories. Interestingly, for both the groups, conspiracy theory related Tweets are Tweeted more than Retweeted. This behavior is contrary to the findings of most studies on Twitter which conclude that users prefer Retweeting to Tweeting \cite{boyd2010tweet}. More Tweeting activity than Retweet activity suggests that although conspiracy related Tweets can be found in climate change discussion, not many users are re-sharing the message. 

\begin{figure}[ht]
    \centering
    \includegraphics[width=0.45\linewidth]{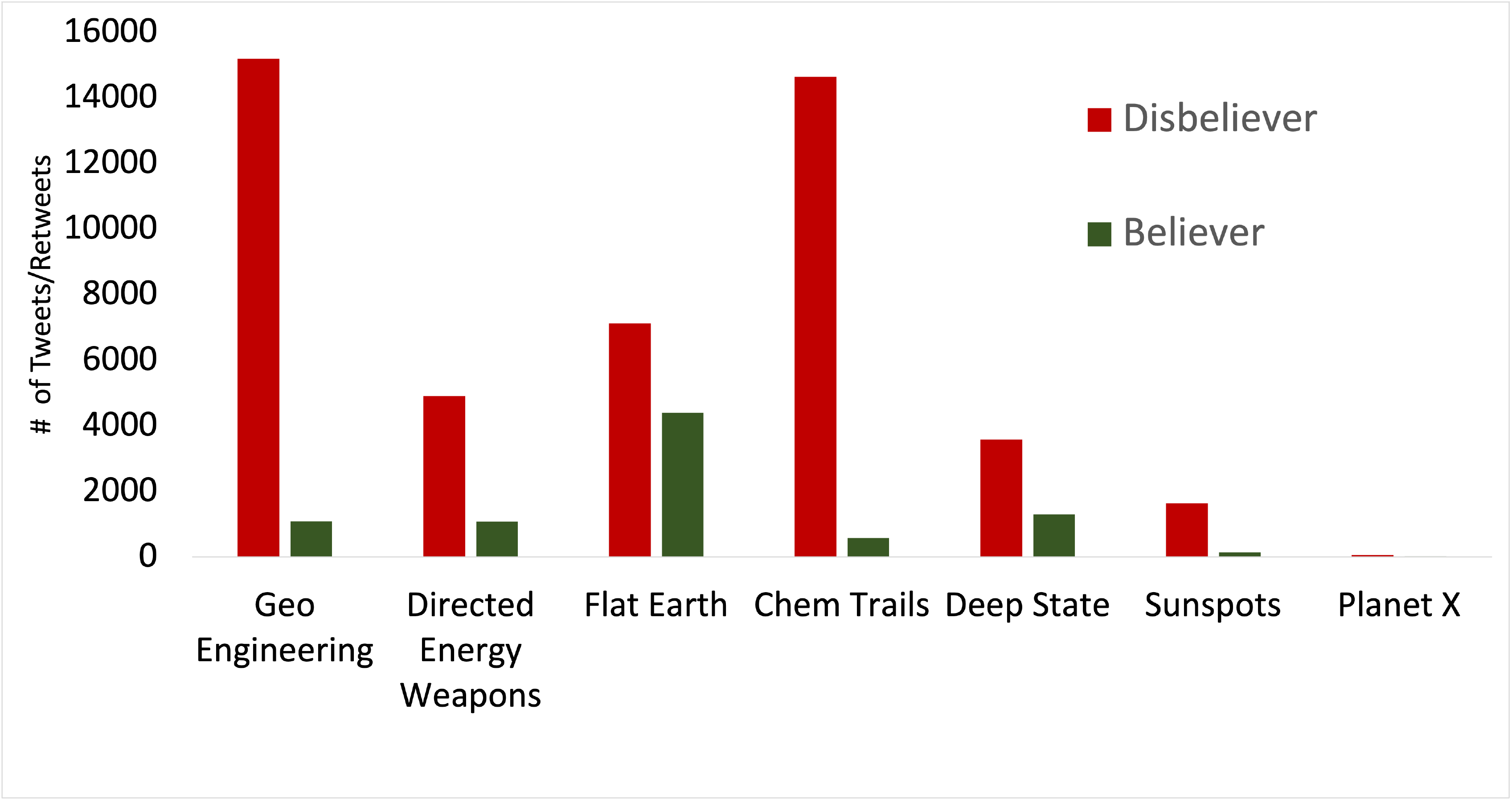}
    \includegraphics[width=0.45\linewidth]{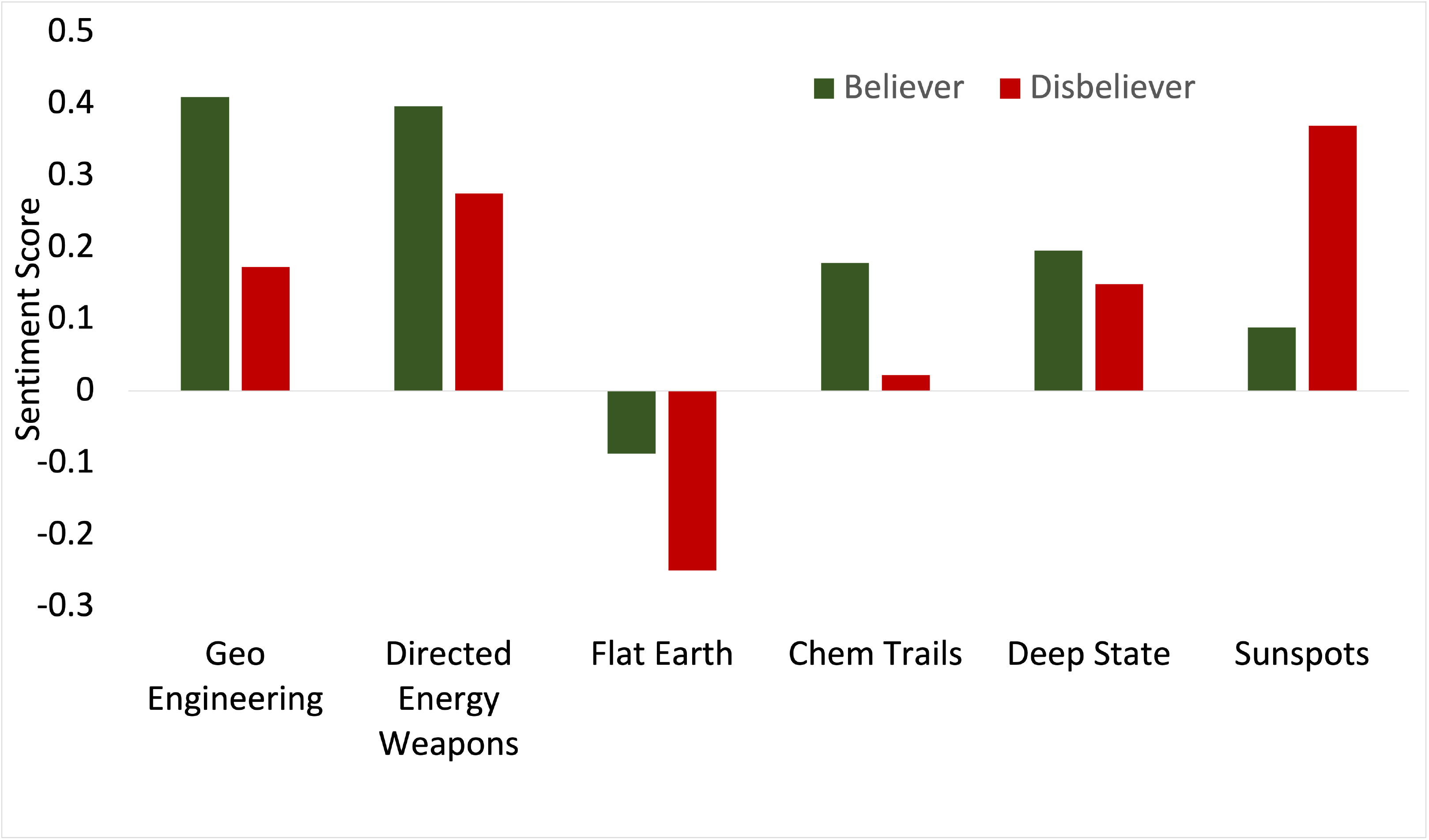}
    \caption{Number of unique Tweets and Retweets shared by Disbelievers and Believers containing different conspiracy theory related keywords defined in \Sref{sec:cons:method} (left).Average sentiment score towards the keywords related to different conspiracy theories. A negative value means a negative sentiment and a positive value means a positive sentiment towards the conspiracy keywords (right).}
    \label{fig:con:conspiracy}
\end{figure}

Once we know that Disbelievers are predominantly sharing the conspiracy theory related Tweets, next, we find which conspiracy theory is most popular. We break down the Tweets/Retweets with the respective type of conspiracy theory in \fref{fig:con:conspiracy}. As expected, Disbelievers are sharing conspiracy theories more than Believers. The most popular conspiracy theory among Disbelievers is Geo-engineering and Chemtrails related conspiracy theory. On the other hand, Believers are sharing Flat Earth conspiracy theory more than other conspiracies. A manual analysis of 100 randomly selected Tweets show that the Flat Earth conspiracy theory is used as a sarcastic comment or to make fun of the other group. We provide further evidence by finding the average sentiment towards conspiracy related keywords \footnote{To find sentiment towards keywords, we utilize Netmapper \cite{carleyora2018} which uses a word-level sentiment computation based on the average of known valences of surrounding words within a sliding window. The output values are between -1 and 1, where negative value represents a negative sentiment, and a positive value represents positive sentiment.}. \Fref{fig:con:conspiracy} reports the average sentiment in Tweets towards conspiracy theory related words. Flat Earth conspiracy theory stands out with negative sentiment, more so when shared by Disbelievers. In other words, irrespective of beliefs about climate change, the Flat Earth conspiracy theory is viewed negatively. Interestingly, Believers have a higher positive sentiment towards ChemTrails and Geo-Engineering conspiracy theories compared to Disbelievers. We suspect that this could be attributed to Believers explaining the actualities of these theories. More robust sentiment analysis with a labeled dataset is needed to draw detailed sentiment level conclusions; such analysis is out of scope for this work.

% {\small

% \begin{table}
% \scriptsize
% \centering
% \caption{Number of unique Tweets and Retweets shared by Disbelievers and Believers containing conspiracy theory related keywords.}
% \begin{tabular}{lrr} 
% \hline
%         & \multicolumn{1}{l}{Disbeliever} & \multicolumn{1}{l}{Believer}  \\ 
% \hline
% Tweet   & 31084                           & 4830                          \\
% Retweet & 14369                           & 3576                          \\
% \hline
% \end{tabular}
% \label{tab:cons:retweet}
% \end{table}
% }

After analyzing the origin of different conspiracy theories, next, we look at the correlation of different conspiracy theories shared by each user. In \tref{tab:cons:corr}, we report the correlation between two different conspiracy theories by finding the number of times different conspiracy keyword is used by each user. We find that most conspiracies are highly correlated with each other, indicating that users who Tweet about one conspiracy also tweet about other conspiracies. The Chemtrails and Unknown Planet conspiracies are least likely to be Tweeted by a user who tweets other conspiracy theories. To further gain insight into the sharing pattern, in \fref{fig:con:conspiracy_dist} we report the number of users sharing unique conspiracy theories. Even on a log scale, we see a steep decline in the number of Believers and Disbelievers sharing different types of conspiracy theories.

Next, we find whether or not the same Tweet has more than one conspiracy discussed. In \tref{tab:tweet:corr}, we report the correlation between two different conspiracy theories by finding the number of times different conspiracy keyword occurs in the same Tweet. We notice in \tref{tab:tweet:corr} that there is a weak negative or close to zero correlation between all the keywords belonging to different conspiracy theories. Twitter users prefer using conspiracy theory keywords independent of using other conspiracy theory keywords in a Tweet. Moreover, conspiracy theories related to Flat Earth and Geo Engineering are most negatively correlated (-0.338). In other words, Twitter users using keywords related to Flat Earth do not use keywords related to Geo Engineering in the same Tweet. Thus, we conclude that most users share one or two types of conspiracy theory and most Tweets have keywords related to one type of conspiracy. Moreover, this behavior does not differ from a change in climate change belief.

\begin{table}
\tiny
\centering
\caption{Correlation matrix of conspiracy theories related keywords used by different users. We find the correlation between two different conspiracy theories by calculating the number of respective keywords used by each user.}
\begin{tabular}{|l|rrrrrrr|} 
\hline
                                                                    & \multicolumn{1}{l}{\begin{tabular}[c]{@{}l@{}}Deep \\State\end{tabular}} & \multicolumn{1}{l}{\begin{tabular}[c]{@{}l@{}}Chem \\Trails\end{tabular}} & \multicolumn{1}{l}{Sunspots} & \multicolumn{1}{l}{\begin{tabular}[c]{@{}l@{}}Directed \\Energy \\Weapons\end{tabular}} & \multicolumn{1}{l}{\begin{tabular}[c]{@{}l@{}}Flat \\Earth\end{tabular}} & \multicolumn{1}{l}{\begin{tabular}[c]{@{}l@{}}Geo \\Engineering\end{tabular}} & \multicolumn{1}{l|}{\begin{tabular}[c]{@{}l@{}}Planet \\X\end{tabular}}  \\ 
\hline
Deep State                                                          & 1.000                                                                    & 0.360                                                                     & 0.713                        & 0.953                                                                                   & 0.958                                                                    & 0.892                                                                         & 0.161                                                                    \\ 
\cline{1-1}
Chem Trails                                                         & 0.360                                                                    & 1.000                                                                     & 0.195                        & 0.333                                                                                   & 0.321                                                                    & 0.361                                                                         & 0.041                                                                    \\ 
\cline{1-1}
Sunspots                                                            & 0.713                                                                    & 0.195                                                                     & 1.000                        & 0.751                                                                                   & 0.740                                                                    & 0.676                                                                         & 0.156                                                                    \\ 
\cline{1-1}
\begin{tabular}[c]{@{}l@{}}Directed \\Energy \\Weapons\end{tabular} & 0.953                                                                    & 0.333                                                                     & 0.751                        & 1.000                                                                                   & 0.982                                                                    & 0.903                                                                         & 0.202                                                                    \\ 
\cline{1-1}
Flat Earth                                                          & 0.958                                                                    & 0.321                                                                     & 0.740                        & 0.982                                                                                   & 1.000                                                                    & 0.915                                                                         & 0.231                                                                    \\ 
\cline{1-1}
\begin{tabular}[c]{@{}l@{}}Geo \\Engineering\end{tabular}           & 0.892                                                                    & 0.361                                                                     & 0.676                        & 0.903                                                                                   & 0.915                                                                    & 1.000                                                                         & 0.184                                                                    \\ 
\cline{1-1}
Planet X                                                            & 0.161                                                                    & 0.041                                                                     & 0.156                        & 0.202                                                                                   & 0.231                                                                    & 0.184                                                                         & 1.000                                                                    \\
\hline
\end{tabular}
\label{tab:cons:corr}
\end{table}

\begin{table}
\tiny
\centering
\caption{Correlation matrix of conspiracy theories related keywords occurring in a single Tweet. We find the correlation between two different conspiracy theories by calculating the number of respective keywords used in each Tweet.}
\begin{tabular}{|l|rrrrrrr|} 
\hline
                                                                    & \multicolumn{1}{l}{\begin{tabular}[c]{@{}l@{}}Deep \\State\end{tabular}} & \multicolumn{1}{l}{\begin{tabular}[c]{@{}l@{}}Chem\\Trails\end{tabular}} & \multicolumn{1}{l}{Sunspots} & \multicolumn{1}{l}{\begin{tabular}[c]{@{}l@{}}Directed\\Energy\\Weapons\end{tabular}} & \multicolumn{1}{l}{\begin{tabular}[c]{@{}l@{}}Flat \\Earth\end{tabular}} & \multicolumn{1}{l}{\begin{tabular}[c]{@{}l@{}}Geo\\Engineering\end{tabular}} & \multicolumn{1}{l|}{\begin{tabular}[c]{@{}l@{}}Planet\\X\end{tabular}}  \\ 
\hline
Deep State                                                          & 1.000                                                                    & -0.179                                                                   & -0.056                       & -0.106                                                                                & -0.151                                                                   & -0.190                                                                       & -0.010                                                                  \\ 
\cline{1-1}
Chem Trails                                                         & -0.179                                                                   & 1.000                                                                    & -0.114                       & -0.218                                                                                & -0.309                                                                   & -0.284                                                                       & -0.021                                                                  \\ 
\cline{1-1}
Sunspots                                                            & -0.056                                                                   & -0.114                                                                   & 1.000                        & -0.065                                                                                & -0.096                                                                   & -0.119                                                                       & -0.006                                                                  \\ 
\cline{1-1}
\begin{tabular}[c]{@{}l@{}}Directed \\Energy \\Weapons\end{tabular} & -0.106                                                                   & -0.218                                                                   & -0.065                       & 1.000                                                                                 & -0.183                                                                   & -0.227                                                                       & -0.012                                                                  \\ 
\cline{1-1}
Flat Earth                                                          & -0.151                                                                   & -0.309                                                                   & -0.096                       & -0.183                                                                                & 1.000                                                                    & \textbf{-0.338}                                                                       & -0.017                                                                  \\ 
\cline{1-1}
\begin{tabular}[c]{@{}l@{}}Geo \\Engineering\end{tabular}           & -0.190                                                                   & -0.284                                                                   & -0.119                       & -0.227                                                                                & \textbf{-0.338}                                                                   & 1.000                                                                        & -0.022                                                                  \\ 
\cline{1-1}
Planet X                                                            & -0.010                                                                   & -0.021                                                                   & -0.006                       & -0.012                                                                                & -0.017                                                                   & -0.022                                                                       & 1.000                                                                   \\
\hline
\end{tabular}
\label{tab:tweet:corr}
\end{table}

% \begin{table}
% \scriptsize
% \centering
% \caption{Fraction of users labeled as bot-like accounts at different probability thresholds.}
% \begin{tabular}{rrr} 
% \hline
% \multicolumn{1}{l}{Threshold} & \multicolumn{1}{l}{Believers} & \multicolumn{1}{l}{Disbeliever}  \\ 
% \hline
% 0.5                           & 0.45                          & 0.46                            \\
% 0.6                           & 0.35                          & 0.36                            \\
% 0.7                           & 0.24                          & 0.27                            \\
% \hline
% \end{tabular}
% \label{tab:botscores}
% \end{table}

% \begin{figure}[ht]
%     \centering
%     \includegraphics[width=0.6\linewidth]{cons_sentiment.pdf}
%     \caption{Average sentiment score towards the keywords related to different conspiracy theories. A negative value means a negative sentiment and a positive value means a positive sentiment towards the conspiracy keywords.}
%     \label{fig:con:conspiracy_sentiment}
% \end{figure}

Lastly, we look at whether bot-like accounts drive the conspiracy theory related discussion. We find that even at 0.7 probability cutoff, about a quarter of all users exhibit bot-like characteristics. We also find that there is not much difference in the activity between Disbelievers and Believers. Moreover, we find that most bots ($\sim$88\%) share only one type of conspiracy theory. This conclusion is similar to the results described in \fref{fig:con:conspiracy_dist}, where we report the distribution without separating bot-like accounts. Moreover, bot-like accounts also show a similar pattern with regards to sharing the type of conspiracy theories. Bot-like accounts showing behavior akin to Disbelievers share more conspiracy theories related to Geo-engineering and Chem Trails. On the other hand, bot-like accounts showing behavior akin to Believers share more conspiracy theories with Flat Earth related keywords. 
{\small
\begin{figure}[ht]
    \centering
    \includegraphics[width=0.5\linewidth]{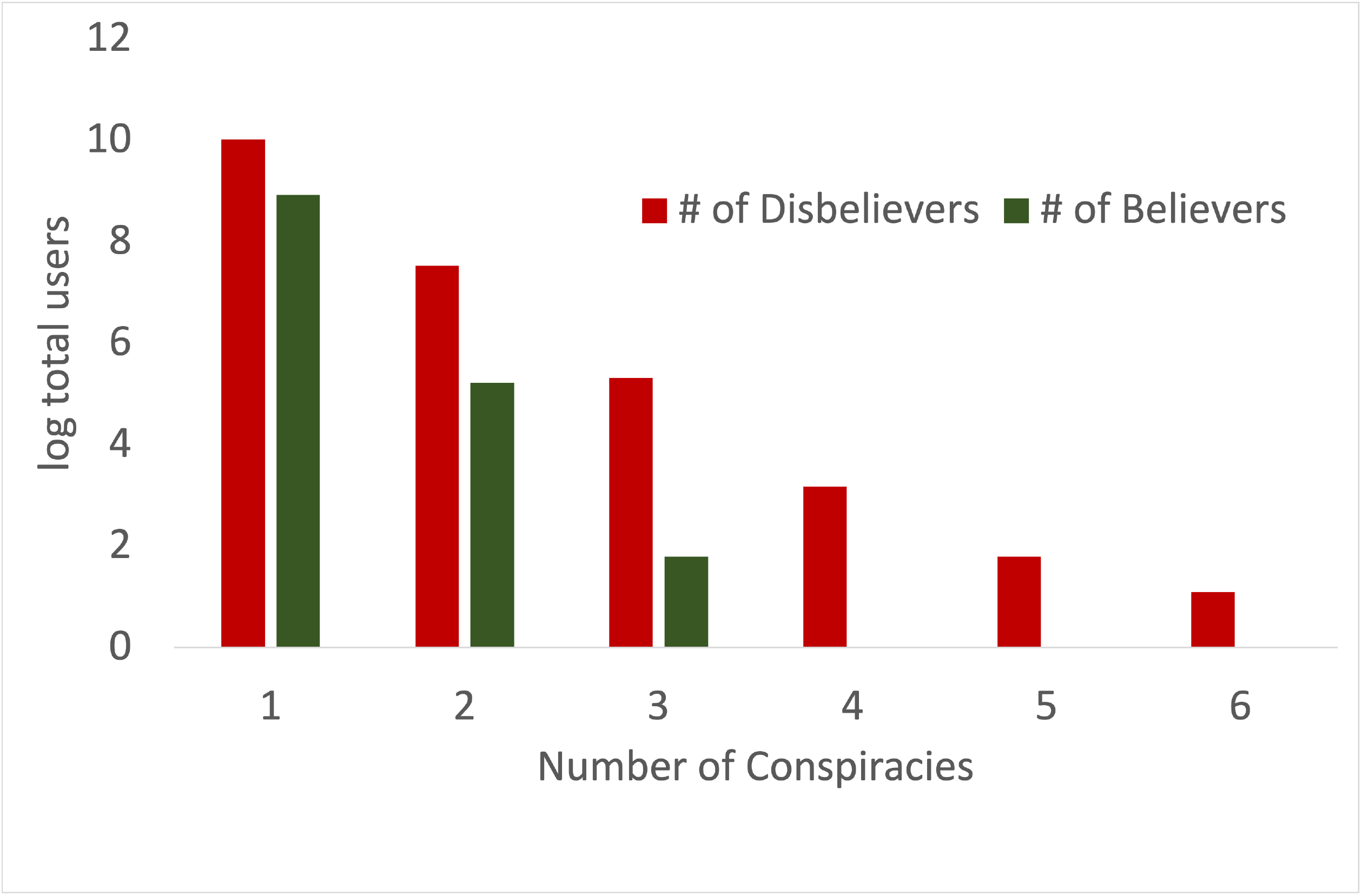}
    \caption{Distribution of Disbeliever and Believer users (in log scale) sharing unique conspiracy theories.}
    \label{fig:con:conspiracy_dist}
\end{figure}
}

\section{Discussion}
\label{sec:cons:discussion}

Understanding people's underlying beliefs helps understand the constructs by which people could be attracted or repelled by different messaging. People believing in conspiracy theories are more likely to believe that a conspiracy theory is a possible explanation of climate change \cite{uscinski2017climate}. Hence, conspiracy theories could be used as a potential recruitment tool by Disbeliever lobbyists. Celebrities and politicians have been vocal about their criticism of science, even using conspiracy theories as possible explanations for climate change \cite{uscinski2017climate}. These reasons make the study of conspiracy theories in the climate change context even more relevant. 

Conspiracy theories are a means for people to justify the actions of a powerful entity or a person, mostly when those actions are not relatable \cite{brotherton2015suspicious,brotherton2013measuring}. \cite{uscinski2017climate} argue that the influence of elites interacting with the masses predispositions explains conspiracy thinking and why there is a partisan divide in such thinking. Moreover, President Donald Trump's election has further enhanced this effect and could potentially lead to mass radicalization \cite{rightwingembrace}. Conspiracy belief is thus linked to people's justification of predisposed climate change belief. Future research on conspiracy theories warrants these explanations to be looked at from the lens of psychology and social science. In this paper, we find that climate change Disbelievers are more likely to share conspiracy theories. The conspiracy theories range from deep state conspiracy theory, which portrays climate change as an agenda of individual actors or \textit{deep state} to possible explanatory theories such as sunspots and chemtrails. Future research should look at these theories from the lens of explanatory or motivated by partisanship. 

Conspiracy theories are a real threat to effective climate change messaging. Climate change messaging should not indiscriminate all conspiracy theories but tackle the popular ones and alienate the unpopular ones. Climate change communication research should look to evolve messaging in ways that take into account different beliefs. Conspiratorial thinking and reasoning to justify climate change will dampen the global effort to decrease climate change effects. In this paper, we show that most people sharing conspiracies in the context of climate change only share one or two types of conspiracies. The most popular conspiracy theories are related to Chem Trails or Geo-Engineering. Policymakers should focus on delivering targeted messages to Disbelievers about the scientific practicalities of these conspiracies. Moreover, climate change Believers using Flat Earth conspiracy theory to target Disbelievers or their belief does not help clear scientific facts. Our results suggest that Flat Earth conspiracy theory is not the most popular conspiracy theory among Disbelievers.  

Previous studies have concluded that Bot-like accounts stir conversations in differently politically aligned belief groups rather than concentrating on conversations in one belief group \cite{tyagi2020brims,bessi2016social}. In this study, we further provide evidence that Bot-like accounts were similarly active in sharing conspiracy related messages irrespective of whether they showed activity akin to a Disbeliever or a Believer. These bot-like accounts aim at widening the divide between belief groups and pose a danger of creating confusion on scientific facts \cite{bessi2016social,broniatowski2018weaponized,ferrara2016rise}. As more and more people consume information via social media, it becomes imperative for these platforms to identify and remove bot-like accounts.   

To the best of our knowledge, this study is the first attempt to find conspiracy theories in climate change in a large social media dataset. We find that some conspiracy theories are more popular and used widely to justify climate change compared to others. Future psychology and social science scholarship should divide conspiratorial thinking into different types of conspiracies. This will help find the underlying constructs and motivations, knowing which helps target climate change communication messaging.

Besides the demographic representativeness of the data, there are other limitations in this analysis. First, although we have many tweets about climate change conspiracies, it does not encompass those interactions that do not include our collection keywords. Second, we use a proxy of keywords to classify Tweets as conspiracy related or not. We do not make an effort to find if sarcasm or negation is used to call out conspiracies; we leave this to future scholarship. Last, we focused on the conspiracy theories recorded in media or found during our search. Many more conspiracy theories could be widespread in the climate change debate. Nevertheless, we believe that we were able to analyze the main conspiracy theories more widely popular among general Twitter users.

\bibliography{references.bib}
\bibliographystyle{splncs04}

\end{document}